\documentclass[useAMS,usenatbib]{mnras}
\usepackage{graphicx,natbib,amssymb,amsmath,stmaryrd,makecell}
\usepackage{txfonts}
\usepackage{xcolor}
\usepackage{hyperref}
\usepackage{xspace}

\newcommand{\Planck}{{\sc Planck}}

\newcommand{\xe}{x_{\rm e}}

\newcommand{\id}{{\,\rm d}}

\newcommand{\beq}{\begin{equation}}   %

\newcommand{\eeq}{\end{equation}}   %

\newcommand{\beqa}{\begin{eqnarray}}   %

\newcommand{\eeqa}{\end{eqnarray}}   %

\newcommand{\beal}{\begin{align}}

\newcommand{\enal}{\end{align}}

\newcommand{\bspl}{\begin{split}}

\newcommand{\espl}{\end{split}}

\newcommand{\bsub}{\begin{subequations}}

\newcommand{\esub}{\end{subequations}}

\newcommand{\bmulti}{\begin{multline}}   %

\newcommand{\beqm}{\begin{mathletters}}   %

\newcommand{\eeqm}{\end{mathletters}}   %

\newcommand{\pot}[2]{#1 \times 10^{#2}}

\newcommand{\COBEF}{{\it COBE/FIRAS}\xspace}
\renewcommand{\Planck}{{\it Planck}\xspace}

\title[Radio signal from accreting PBHs]{Can accreting primordial black holes explain the excess radio background?}

\author[Acharya et al.]
{Sandeep Kumar Acharya$^1$\thanks{E-mail:sandeep.acharya@manchester.ac.uk}, 
Jiten Dhandha$^1$
and 
Jens Chluba$^1$
\\
$^1$Jodrell Bank Centre for Astrophysics, School of Physics and Astronomy, The University of Manchester, Manchester M13 9PL, U.K.
}

\begin{document}

\maketitle

\begin{abstract}
The excess radio background seen at $\simeq 0.1-10\,{\rm GHz}$ has stimulated much scientific debate in the past years. 
Recently, it was pointed out that the soft photon emission from accreting primordial black holes may be able to explain this signal. We show that the expected ultraviolet photon emission from these accreting black holes would ionize the universe completely at $z>6$ and thus wash out the 21 cm absorption signature at $z\simeq$ 20 as well as be in tension with existing cosmic microwave background anisotropy and average spectral distortion limits. We discuss possible augmentations of the model; however, it seems that an explanation of radio excess by accreting primordial black holes is not well-justified.
\end{abstract}

\section{Introduction}

The detection of a bright radio monopole at $\simeq 0.1-10$ GHz \citep{Arcade2011,DT2018} with an intensity larger than the standard Cosmic Microwave Background (CMB) and known galactic and extragalactic radio sources \citep{PB1996} is one of the outstanding problems in current astrophysics and cosmology. On top of this, the possible detection of an unexpectedly-strong 21 cm absorption feature at $z\simeq 20$ \citep{Edges2018} further adds to this deepening mystery, although recent independent efforts to verify this measurement have resulted in a null detection \citep{Saras2022}. The reader is referred to \citet{Singal2018} for a detailed review of possible explanations for radio excess and future experimental efforts to verify its detection. Most explanations invoke unresolved extragalactic radio contributions, but also galactic contribution could be relevant \citep{SC2013}.

One possible cosmological explanation of the radio excess could be dark matter halos hosting unresolved radio sources \citep{H2014}. This naturally implies that the background is anisotropic, as the halos are clustered and have finite sizes. However, it was shown that the expected anisotropic signal violates the observed limits unless the radio emitting sources are extended over $\gtrsim {\rm Mpc}$ scale and are located at redshifts $z>5$ \citep{H2014}. 

Other explanations for radio excess include Comptonized photon injection distortions \citep{Chluba2015GreensII, Bolliet2020PI}, annihilating axion-like dark matter \citep{Fraser2018}, dark photons \citep{PPRU2018, Caputo2022}, supernova explosion of population III stars \citep{JNB2019}, superconducting cosmic strings \citep{BCS2019}, decay of relic neutrinos to sterile neutrinos \citep{CDFS2018}, thermal emission of quark nugget dark matter \citep{LZ2019}, bright luminous galaxies \citep{MF2019} and accreting astrophysical black holes \citep{ECLDSM2018,ECL2020}. 
Many of these works were stimulated by the EDGES measurements and thus simultaneously attempt to explain the large 21 cm absorption feature. 

One interesting possibility for testing the presence of an extragalactic radio background proposes to use the up-scattering of the photon field when crossing clusters of galaxies \citep{HC2021, Lee2022rSZ}. This effect, coined radio-SZ effect, is the radio background analog of the Sunyaev-Zeldovich effect \citep{Zeldovich1969}, and might be observable at $\lesssim 3\,{\rm GHz}$ using experiments such as MeerKAT or LOFAR. Along another route, the interaction of cosmic rays with the radio background could produce ultra-high energy photons from pair-production that could test the origin of the radio excess \citep{GKS2022}.  

Recently, \cite{MK2021} studied the possibility of explaining the radio monopole with a population of accreting supermassive primordial black holes (PBH) with masses $M=10^5-10^{12}M_{\odot}$. In this, it is crucial to have a relation between the radio and X-ray luminosity of the PBHs.
Previous works \citep{ROM2008,AK2017,HHMRVV2018,MPVW2019,Y2021} have studied accretion onto PBHs (though they restrict themselves to $M\lesssim 10^4 M_{\odot}$) and derived constraints on their abundance using the CMB anisotropy and global 21cm signature. These limits were derived by deducing the X-ray luminosity under the assumptions that the accretion process is spherical, which adds some level of uncertainty. 
Instead, the authors in \cite{MK2021} use the parametric relation between radio and X-ray luminosity of \cite{Lusso2010} observed at low redshifts. Using well motivated astrophysical parameters and a reasonable estimate of PBH abundance, it was shown in \cite{MK2021} that radio emission from these PBHs could explain the radio excess of \cite{Arcade2011,DT2018}. 

Accretion onto black holes will not only result in radio and X-ray emission but also UV/optical emission \citep{Shakura1973}. Energetic photons with $E>13.6$~eV can ionize neutral hydrogen and heat the baryonic gas. Therefore, one has to study the possibility that the over-abundance of ionizing photons can reionize the universe much before $z\simeq 6$. This would also heat the gas sufficiently such that we may not see any 21 cm absorption signature, which can be computed using standard methods \citep{PL2012}. 

In this paper, we use a parametric relation between UV and X-ray luminosity \citep{Lusso2015} similar to the relation used in \cite{MK2021} to study the effect of UV emission from accreting PBHs on the ionization and thermal history of the Universe. We show that the fiducial parameters used in \cite{MK2021} predict too many energetic photons which can ionize the Universe before $z\simeq 6$ and thereby wash out any 21 cm absorption signal at $z\simeq 20$. In addition, the changes to the ionization history induce modifications to the CMB temperature and polarization anisotropies that are ruled out by \Planck \citep{Planck2015params}.
Finally, we highlight that a large part to the required PBH population has already been ruled out by \COBEF.
Therefore, an explanation of the radio excess by accreting PBHs alone seems to run into significant constraints. Using a simple conservative estimate, we show that the abundance of PBHs has to be orders of magnitude smaller than what has been assumed in \cite{MK2021}. 


\section{Photon emission from accreting PBHs}
\label{sec:photon emission}

\subsection{Radio background from PBHs}


In this section, we briefly discuss the calculation of radio background from accreting primordial black holes, following \cite{MK2021}. The readers are referred to this work for a more detailed discussion.
Accretion of matter onto supermassive black holes results in non-thermal photon emission in radio, UV/optical and X-ray. 
To compute the radio luminosity at 1.4~GHz, we use the luminosity relation of \cite{WWK2006}, which was also applied by the authors in \cite{MK2021}. The radio luminosity of each black hole at 1.4~GHz is then related to its bolometric X-ray luminosity in the 0.1-2.4 keV band by \citep{WWK2006},
\begin{equation}
    \left.\log_{10}(L_R/L_E)\right|_{\rm 1.4GHz}=\left.0.86\log_{10}(L_X/L_E)\right|_{\rm 0.1-2.4 keV}-5.08,
\end{equation}
which can be cast into the form $\left(L_R/L_E\right)\approx\left(L_X/L_E\right)^{0.86}\times 10^{-5}$ with $L_E=1.26\times 10^{31} (M/M_{\odot})$~W. From here onward, we drop the energy (or frequency band)  subscript from the luminosity for brevity and simply refer to radio/X-ray luminosity as $L_R$ or $L_X$. The X-ray luminosity is related to the Eddington luminosity ($L_E$) as,
\begin{equation}
L_X=f_{{\rm BH},X}\lambda L_E.
\end{equation}
The ratio of the bolometric luminosity to the Eddington luminosity is given by $\lambda$, while $f_{{\rm BH},X}$ is the ratio of $L_X$ to bolometric luminosity. We choose the values $f_{{\rm BH},X}=\lambda=0.1$ which was used as the fiducial values by the authors in \cite{MK2021}.

The specific luminosity in the vicinity of $\nu=1.4$ GHz is given by,
\begin{equation}
    l_R(\nu)=\left(\frac{\nu}{1.4 {\rm GHz}}\right)^{-0.6}l_R(\nu=1.4 {\rm GHz}),
\end{equation}
where $\nu l_R(\nu)=L_R$. The comoving radio emissivity is given by,
\begin{equation}
    \epsilon(\nu)=n_{\rm PBH}l_R(\nu),
\end{equation}
where $n_{\rm PBH}$ is the number density of primordial black holes which is explicitly written as,
\begin{equation}
    n_{\rm PBH}=\frac{f_{\rm PBH}\,\rho_{\rm DM}}{M_{\rm PBH}}.
\end{equation}
 The fractional energy density of primordial black holes compared to total dark matter density $\rho_{\rm DM}$ is given by $f_{\rm PBH}$  and $M_{\rm PBH}$ is the mass of black holes. The authors in \cite{MK2021} assumed the mass of the black holes to be in the range $10^5-10^{12}M_{\odot}$.  

The comoving radio emissivity from PBHs per unit energy is then given by \citep{MK2021},
\begin{equation}
    \epsilon(E)=5.65\times 10^{19}f_{\rm duty}(f_{{\rm BH},X}\lambda)^{0.86}\left(\frac{f_{\rm PBH}\,\rho_{\rm DM}}{1\rm{kgm^{-3}}}\right)\left(\frac{E}{5.79\mu \rm{eV}}\right)^{-0.86} \rm{m^{-3}s^{-1}}.
\end{equation}
where $f_{\rm duty}$ is the duty cycle and specific luminosity is per unit energy basis. The fiducial values of $f_{\rm duty}=0.01$ and $f_{\rm PBH}=10^{-4}$ [the maximal value of the halo dynamical friction constraint quoted in mass range $10^5-10^{12}M_{\odot}$ \citep{CK2020}], which were the values chosen by the authors. Note that $M_{\rm PBH}$ drops out in the emissivity, but the astrophysical parameters $f_{\rm duty}, \lambda, f_{{\rm BH},X}$ have implicit mass dependence. This technical detail will not be important for the point we are trying to make. The specific intensity of radio background at $E$ is then given by \citep{ECLDSM2018},
\begin{equation}
    J(E,z)=\frac{c}{4\pi}(1+z)^3\int_z ^{\infty} \frac{\epsilon(E')}{1+z'}\frac{{\rm d}z'}{H(z')},
    \label{eq:intensity}
\end{equation}
where $E'=E(1+z')/(1+z)$. In the Rayleigh-Jeans regime, the temperature due to this radio background is given by,
\begin{equation}
    T(E)=\frac{h^3c^2}{2k_B}\frac{J(E)}{E^2},
    \label{eq:temperature}
\end{equation}
The temperature at $E=5.79\mu$eV ($\nu=1.4$ GHz) at $z=0$ turns out to be $\simeq 0.02$~K \citep{MK2021}. 

\begin{figure}
\centering 
\includegraphics[width=0.95\columnwidth]{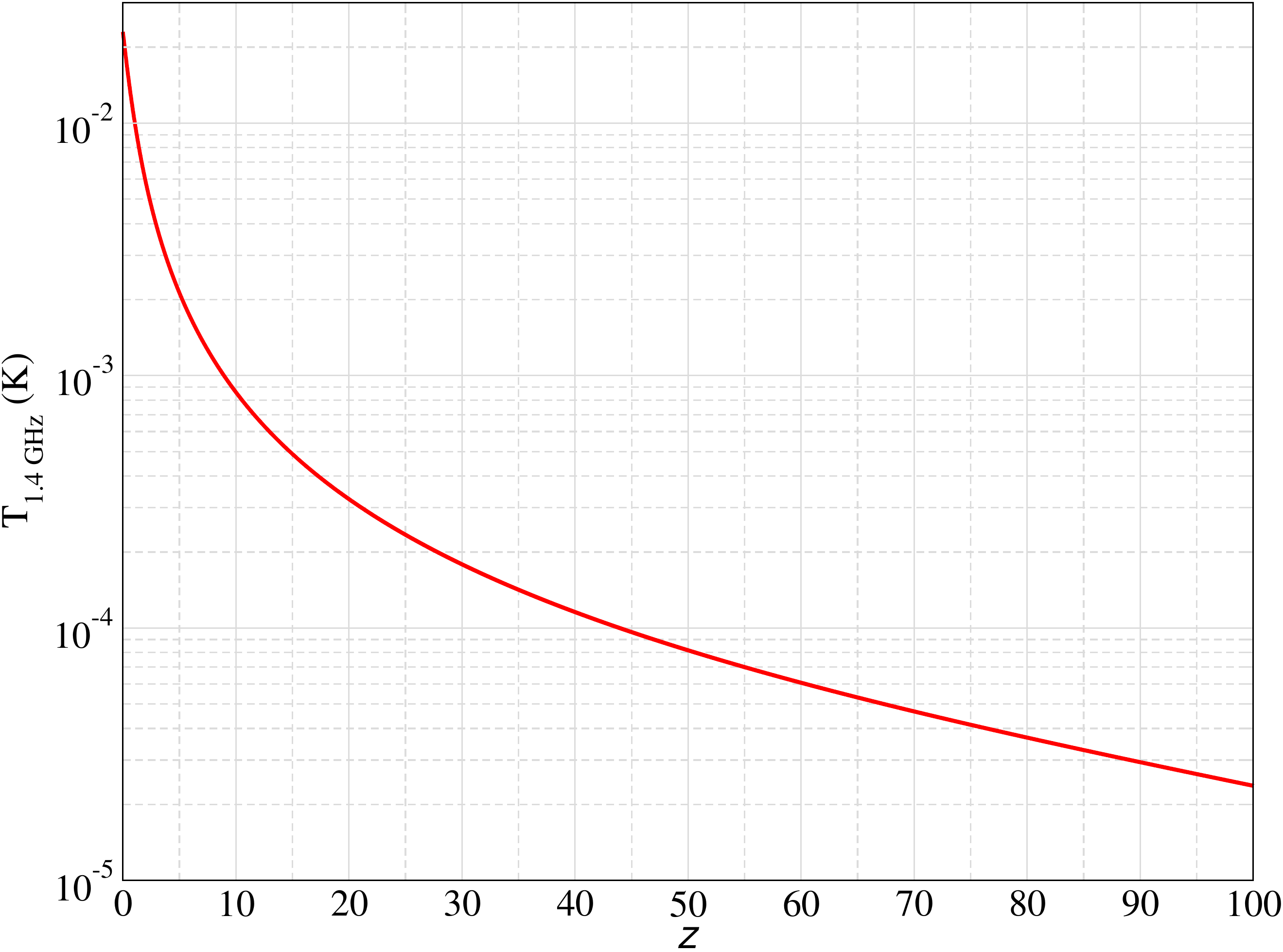}
\\
\caption{Comoving intensity of radio background as a function of redshift at 1.4 GHz as the PBHs accrete and the background builds up. We have converted the comoving intensity to temperature using Eq.~\eqref{eq:temperature}. }
\label{fig:radio_background}
\end{figure}

\begin{figure}
\centering 
\includegraphics[width=0.95\columnwidth]{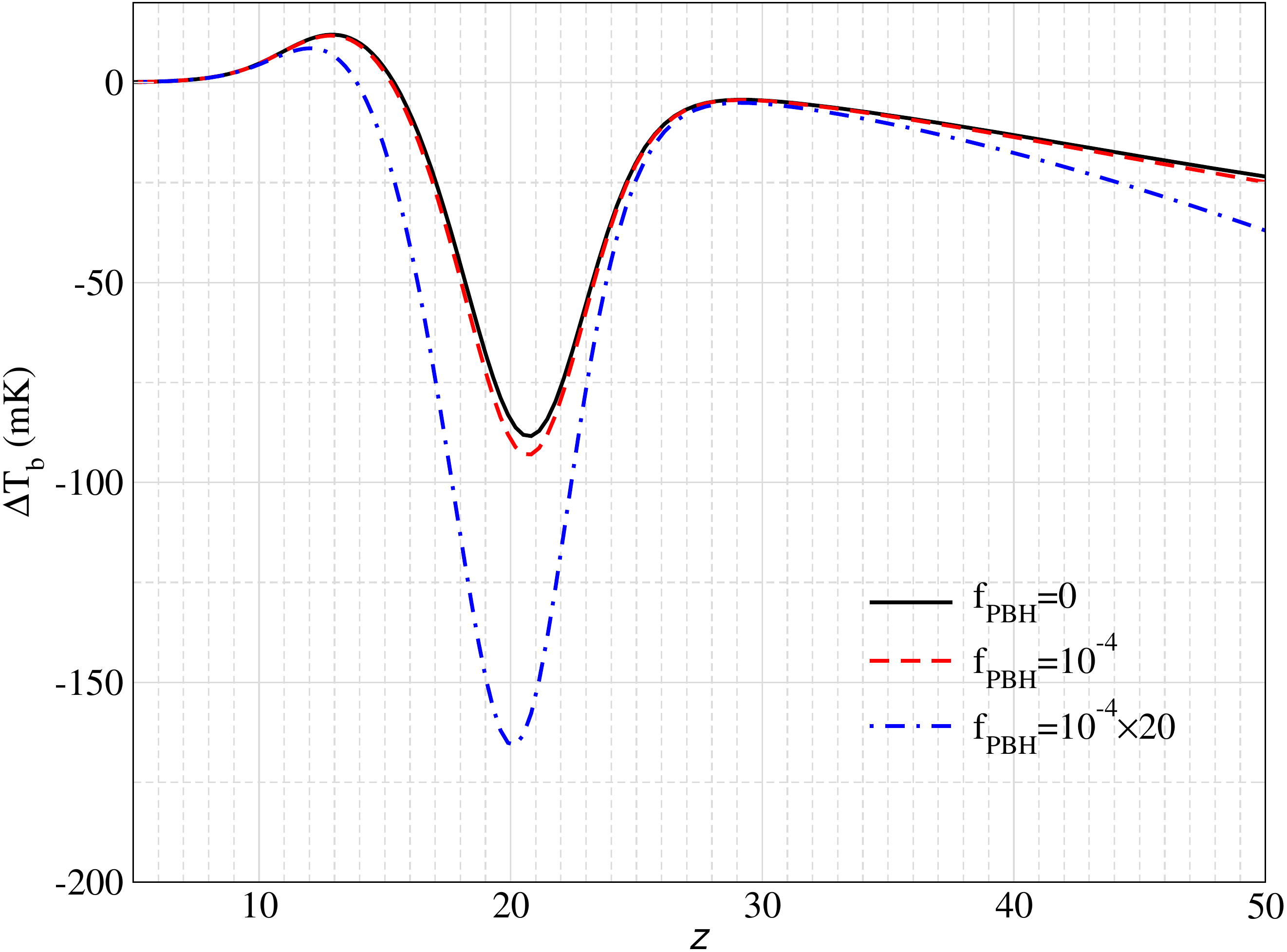}
\\
\caption{The 21cm distortion as a function of redshift for various values of $f_{\rm PBH}$. The fiducial parameters are $f_{{\rm BH},X}=\lambda=0.1$ and $f_{\rm duty}=0.01$. We have ignored any modification to thermal history of Universe such as changes in electron fraction and matter temperature. We also show the case where we fix $f_{\rm PBH}$ to $10^{-4}$ but tune $f_{{\rm BH},X}, \lambda$ and $f_{\rm duty}$ such that radio luminosity is boosted by a factor of 20. With a further boost of a factor 2-3, the parameters could help explain the ARCADE excess.}
\label{fig:21cm_abs_signal}
\end{figure}

We illustrate the buildup of radio background as a function of redshift in Fig. \ref{fig:radio_background}. The authors in \cite{MK2021} argue that by choosing $f_{\rm duty}, \lambda, f_{{\rm BH},X}$ and keeping $f_{\rm PBH}$ unchanged, it is possible to make $\left.T\right|_{\rm 1.4 GHz}\simeq 0.5$~K, which can explain the radio excess of \cite{Arcade2011,DT2018}. 
In Fig.~\ref{fig:21cm_abs_signal}, we show the expected 21cm absorption signal from accreting PBHs (see Sect.~\ref{sec:21cmmodel} for details of the 21cm modeling). One can see that  with a further boost factor of 2-3, it may indeed be possible to explain EDGES result \citep{Edges2018}. 
However, we are going to show that the associated UV photon emission, assuming the fiducial values, is enough to ionize the Universe much before reionization. This effect was ignored in the computation of the 21cm signal but does modify it significantly. 
%

\subsection{UV luminosity from PBHs}
\begin{figure}
\centering 
\includegraphics[width=0.95\columnwidth]{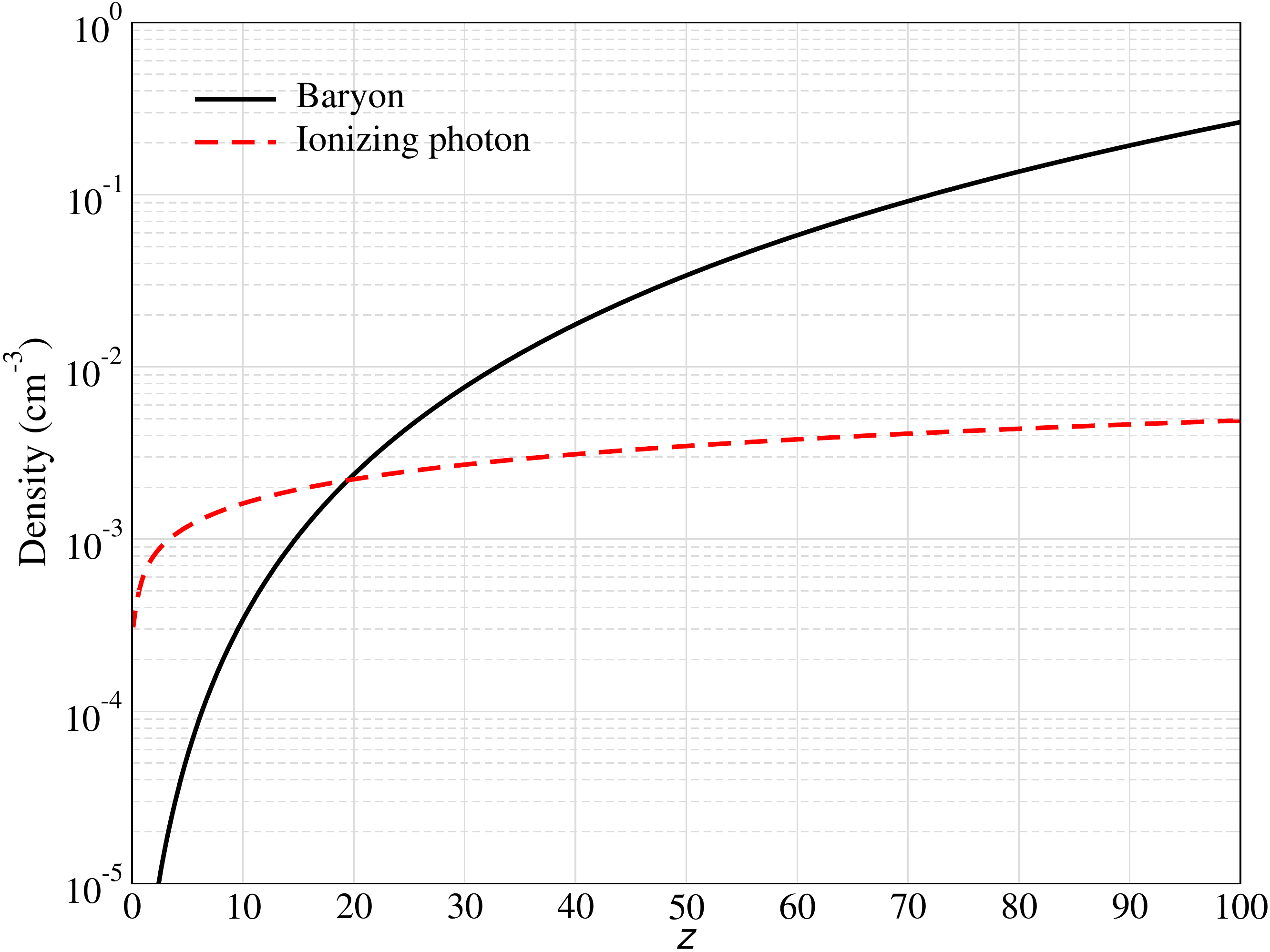}
\caption{Baryon density (in black) and ionizing photon number density (red) which is $\dot{N}(z)\Delta t$ with $\Delta t=\frac{1}{(1+z)H(z)}$. The parameters used are for $f_{\rm PBH}=10^{-4}$ with $f_{{\rm BH},X}=\lambda=0.1$, $f_{\rm duty}=0.01$. }
\label{fig:density}
\vspace{-3mm}
\end{figure}

As explained in the previous section, the X-ray luminosity of black holes is given by $L_X=f_{\rm duty}f_{{\rm BH},X}\lambda L_E$, which alternatively can be written as,
\begin{equation}
    L_X=f_{\rm duty}f_{{\rm BH},X}\lambda\times 7.86\times  10^{49} \frac{M}{M_{\odot}} {\rm eV s^{-1}}.
 \end{equation}
For X-ray luminosity, we have included the $f_{\rm duty}$ factor just like radio luminosity. The luminosity at 2500 $\AA$ ($\approx$ 5eV) is related to the luminosity at 2 keV by the relation \citep{Lusso2010},
\begin{equation}
    \log_{10}\left(\frac{L_{2 \rm keV}}{L_{2500 \AA}}\right)\approx -1.37 \,\log_{10}\left(\frac{\nu_{2 \rm keV}}{\nu_{2500 \AA}}\right)\approx -2.605\times 1.37,
\end{equation}
which implies\footnote{We would like to clarify that $\nu_{2 {\rm keV}}$ and $\nu_{2500\AA}$ corresponds to the frequency at the corresponding energy or wavelength and there is no additional conversion factor involved. Same applies for $L_{2 \rm keV}$ and $L_{2500 \AA}$ which is the luminosity per frequency $\nu$ corresponding to 2keV/2500$\AA$.} $L_{2500 \AA}=L_{\rm 2 keV}\times 10^{3.569}$. 

We next have to convert the bolometric luminosity in 0.1-2.4 keV band to the luminosity at 2 keV. For this, we write the bolometric luminosity as,
\begin{equation}
    L_X=\int_{0.1 {\rm keV}}^{2.4 {\rm keV}}L_{0.1 {\rm keV}}\left(\frac{\nu}{0.1 {\rm keV}}\right)^{1-\Gamma_X} {\rm d}\nu.
\end{equation}
The value of $\Gamma_X\approx 2$ \citep{Lusso2015,Sacchi2022}; however, the results do not crucially depend on this. Then $L_{{\rm 2 keV}}$ is given by,
\begin{equation}
    L_{{\rm 2 keV}}=\frac{L_X}{0.1 {\rm keV}\times 3.18}\left(\frac{2}{0.1}\right)^{-1}.
\end{equation}
We can extrapolate the luminosity at 912 $\AA$ ($\approx$ 13.6 eV) from luminosity at 2500 $\AA$ using the power law \citep{Lusso2015},
\begin{equation}
    L_{912\AA}=L_{2500\AA}\left(\frac{13.6}{5}\right)^{-0.65}
\end{equation}
This makes the luminosity at 912 $\AA$ smaller than 2500 $\AA$ by a factor of $\approx 2$. Blueward of 912 $\AA$, the power law index of luminosity is -1.7 \citep{Lusso2015}. The total UV luminosity is given by,
\begin{equation}
    L_{\rm UV}=\frac{L_{2500\AA}}{2}\int_{13.6 {\rm eV}}^{\infty}\left(\frac{\nu}{13.6 {\rm eV}}\right)^{-1.7}{\rm d}\nu.
\end{equation}
Since, the spectrum is very steep, the total luminosity is dominated by the lower limit and we ignore contribution from higher energy. Putting everything together, we have,
\begin{equation}
    L_{\rm UV}=\frac{1}{2\times 0.7}\times \frac{13.6 {\rm eV}}{0.1 {\rm keV}\times 3.18}\times \frac{1}{20}\times L_X\times 10^{3.569},
\end{equation}
which can be simplified to, 
\begin{equation}
    L_{\rm UV}\approx 5.66L_X.
\end{equation}
We will use this simple relation in our computation below.

\subsection{Global 21-cm signal modelling}
\label{sec:21cmmodel}
In this section, we briefly discuss the modelling of the global 21-cm cosmological signal. The 21-cm brightness measured against the background CMB temperature, at redshift $z$, is given by its differential brightness temperature \citep[see e.g.][]{FOB2006},
\begin{equation}
\label{eq:DTb}
    \Delta T_\text{b} = \dfrac{\left(1-{\rm e}^{-\tau_{21}}\right)}{1+z}\left(T_\text{s}-T_\text{CMB} \right)
\end{equation}
where $T_\text{CMB}$ is the CMB temperature, $T_\text{s}$ is the 21-cm spin temperature and $\tau_{21}$ is the 21-cm optical depth. The signal is observed in absorption when $\Delta T_\text{b} <0$ and emission when $\Delta T_\text{b} >0$. The spin temperature in turn is defined by the ratio of the population of the upper and lower hyperfine states:
\begin{equation}
\dfrac{n_1}{n_0} \equiv 3 {\rm e}^{-T_\star/T_\text{s}}
\end{equation}
where $T_\star = h\nu_{21}/k_\text{B}=0.068\text{~K}$, $\nu_{21} = 1.42\text{GHz}$, $h$ is the Planck's constant, $k_\text{B}$ is the Boltzmann constant and the constant $3$ is a statistical degeneracy factor. 

The modelling of the spin temperature is done in a similar fashion to \citet{Mittal2022} with a few minor modifications. 
For any model of 21-cm signal, three main factors affect the evolution of the spin-temperature with redshift: interaction of HI with CMB photons (radiative coupling); collisions with other hydrogen atoms, electrons and protons (collisional coupling); and the resonant scattering of UV photons from stars as they redshift into Lyman-$\alpha$ line (Wouthuysen-Field coupling, shortened as WF coupling). As defined in \citet{Venu2018}, the spin temperature is given by:
\begin{equation}
T_\text{s} ^{-1}= \dfrac{x_\text{CMB} T_\text{CMB}^{-1} + x_{\rm c} T_\text{M}^{-1} + x_\alpha T_\alpha^{-1}}{x_\text{CMB}+ x_{\rm c} +x_\alpha}
\end{equation}
where $x_\gamma$, $x_{\rm c}$ and $x_\alpha$ are the radiative, collisional and WF coupling coefficients respectively, $T_\text{M}$ is the matter temperature and $T_\alpha$ is the colour temperature of the Ly$\alpha$ radiation field. The radiative coupling coefficient $x_\text{CMB}$ itself depends on the spin temperature through the 21-cm optical depth $\tau_{21}$, and is solved iteratively as done in \citet{Mittal2022}, following \citet{Fialkov2019}. The effect of the Ly$\alpha$ background is often approximated in literature by assuming $T_\alpha \simeq T_\text{M}$ \citep[e.g.][]{Mittal2022} or other means \citep[see review][]{Barkana2016}, but can be incorporated accurately by the complex fitting and iterative formalism introduced in \citet{Hirata2006}, which we adopt here. The Ly$\alpha$ spectral energy distribution (SED) $\phi_\alpha(E)$ is assumed to be a broken power-law with a Pop II base model having spectral index of $\alpha_S= 0.14$ \citep{Barkana2005}. The comoving Ly$\alpha$ emissivity is then
\begin{equation}
    \epsilon_\alpha(E,z) = f_\alpha \phi_\alpha(E) \dfrac{\dot{\rho}_\star(z)}{m_\text{b}}
\end{equation}
where $f_\alpha$ is a scaling factor for strength of Ly$\alpha$ background, $\dot{\rho}_\star$ is the comoving star formation rate density (SFRD) and $m_\text{b}$ is the number-averaged baryon
mass. Note that this only accounts for the stellar contribution to the Ly$\alpha$ background; the contribution from accreting PBHs has been neglected, which could induce the WF coupling much earlier. The effect is negligible within the constraints derived in this work, as we shall see in Sect. \ref{sec:thermal_History}.

Finally, the collisional coupling coefficient $x_{\rm c}$ can be calculated using the recombination history from {\tt Recfast++} \citep{CT2011} and spin-exchange rate coefficients tabulated in literature: \citet{Zyg2005} and \citet{FOB2006} for neutral hydrogen collisions; \citet{FF2007a} for electron-hydrogen collisions; and \citet{FF2007b} for proton-hydrogen collisions. The rates are interpolated for intermediate values of $T_\text{M}$ on a log-log scale.

To also account for the possible modifications of the background photon field at low frequencies due to the PBHs emission, in Eq.~\eqref{eq:DTb} we replace $T_{\rm CMB}$ with the brightness temperature evaluated at the 21cm rest frame frequency (see Fig.~\ref{fig:radio_background}). This allows us to estimate the net 21cm signal with respect to the enhanced background.

\vspace{-3mm}
\subsection{Reionization model}
A simple reionization treatment is included, following the work of \citet{Furlanetto2006}, with the ionization rate connected to star-formation history through
\begin{subequations}
\begin{align}
    \dfrac{\id x_{\rm HII}}{\id t} &= \xi_\text{ion}(z)(1-x_{\rm HII})\dfrac{\id f_\text{coll}}{\id t} - \alpha_A C\,x_{\rm HII} \,n_{\rm e} 
    \\
    \dfrac{\id x_{\rm HeII}}{\id t} &= \xi_\text{ion}(z)(f_\text{He}-x_{\rm HeII})\dfrac{\id f_\text{coll}}{\id t} - \alpha_A C\,x_{\rm HeII} \,n_{\rm e},
\end{align}
\end{subequations}
where $x_{\rm HII}=\frac{n_{\rm HII}}{n_{\rm H}}$, $x_{\rm HeII}=\frac{n_{\rm HeII}}{n_{\rm H}}$ are ionized fractions, $f_\text{He}=\frac{n_{\rm He}}{n_{\rm H}}$ is the fraction of helium, $\xi_\text{ion}$ is the ionizing efficiency parameter, $f_\text{coll}$ is the fraction of matter in collapsed dark-matter (DM) haloes with mass $m > m_\text{min}$, $\alpha_A$ is the case-A recombination coefficient, $C \equiv \langle n_{\rm e}^2\rangle/\langle n_{\rm e}\rangle^2$ is the clumping factor, and $n_{\rm e}$ is the total electron number density. Note that an extra factor of $(1-x_{\rm HII})$ and $(f_\text{He}-x_{\rm HeII})$ has been included compared to \citet{Furlanetto2006}; this is physically motivated and takes into account that no ionization can occur without a target. 

The ionizing efficiency parameter is given by
\begin{equation}
    \xi_\text{ion} = A_\text{He} f_\star f_\text{esc}N_\text{ion}
\end{equation}
where $A_\text{He}$ is a correction factor for presence of helium, $N_\text{ion}$ is the number of ionizing photons per baryon, $f_\text{esc}$ is the fraction of ionizing photons escaping host halo and $f_\star$ is star formation efficiency. The collapse matter fraction $f_\text{coll}$ is modelled using the Press-Schechter formalism \citep{PS1974} calculated using {\tt COLOSSUS} \citep{COLOSSUS}, where $m_\text{min}$ varies with redshift assuming a minimum virial temperature for star formation of $T_\text{vir} = 10^4\text{K}$ \citep{DF2018}. The complex effects of the clumping factor are difficult to capture, but a simple analytic expression presented in \citet{Mellema2006} provides a reasonable reionization history. For calculating the Ly$\alpha$ background produced by stars, we need the SFRD as described in the previous section. This is given by
\begin{equation}
    \dot{\rho}_\star(z) = f_\star \bar{\rho}^0_\text{b}\dfrac{\id f_\text{coll}}{\id t}
\end{equation}
where $\bar{\rho}^0_\text{b}$ is the mean cosmic baryon mass density measured today \citep{Furlanetto2006}.

The X-ray heating of electrons is treated as in \citet{Furlanetto2006}, 
\begin{equation}
    \dfrac{2}{3}\dfrac{\epsilon_X}{k_\text{B}nH(z)} = 10^3\text{K}~f_X \left(\dfrac{f_\star}{0.1}\dfrac{f_{X,h}}{0.2}\dfrac{\id f_\text{coll}/\id z}{0.01}\dfrac{1+z}{10}\right)
\end{equation}
where $f_{X,h} \simeq (1+2x_{\rm e})/3$ is the fraction of X-ray energy into heating \citep{Chen2004}, and $f_X$ is a scaling factor for strength of X-ray emissions. The effects of Ly$\alpha$ heating and radiative heating \citep[see][]{Venu2018} are omitted in this work.

For our purposes, we use a fiducial set of parameters ($N_\text{ion}$, $f_\star$, $f_\text{esc}$, $f_\alpha$, $f_X$) = (40000, 0.1, 0.1, 1.0, 1.0). All values but the one for $N_\text{ion}$ are commonly used in the literature \citep[see, e.g.,][]{FOB2006,PL2012,Cohen2017}. For $N_\text{ion}$, a higher value was adopted since, with the (physically-motivated) extra factors of $(1-x_{\rm HII})$ and $(f_\text{He}-x_{\rm HeII})$,  reionization process did not complete at $z\lesssim 6$ in the standard scenario. However, this part of the evolution does not affect the results discussed here significantly.


\section{Modification to ionization and thermal history of universe}
\label{sec:thermal_History}

\begin{figure}
\centering 
\includegraphics[width=0.95\columnwidth]{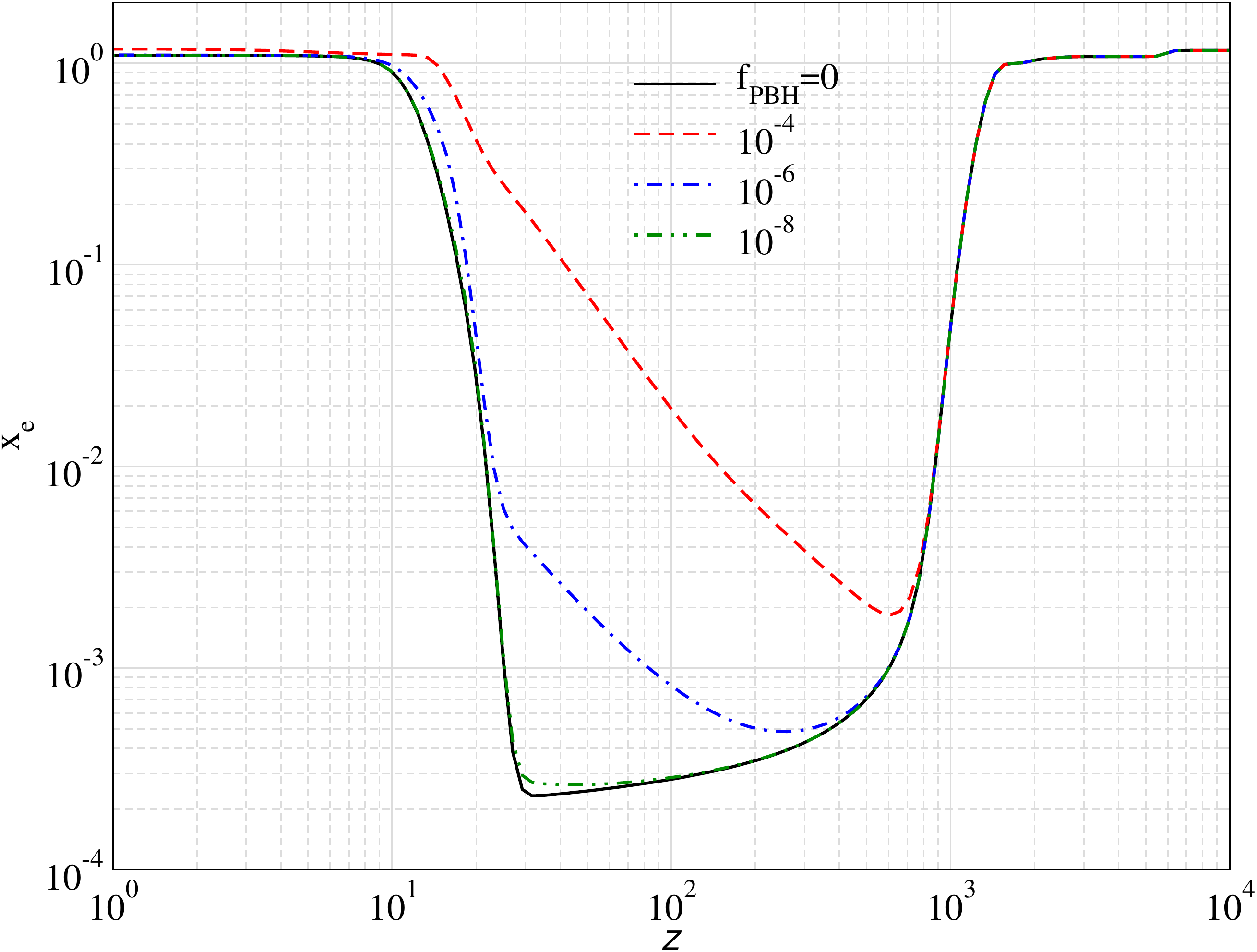}
\\
\caption{Evolution of $x_{\rm e}$ as a function of redshift for a few different $f_{\rm PBH}$. }
\label{fig:xe}
\end{figure}


\begin{figure}
\centering 
\includegraphics[width=0.95\columnwidth]{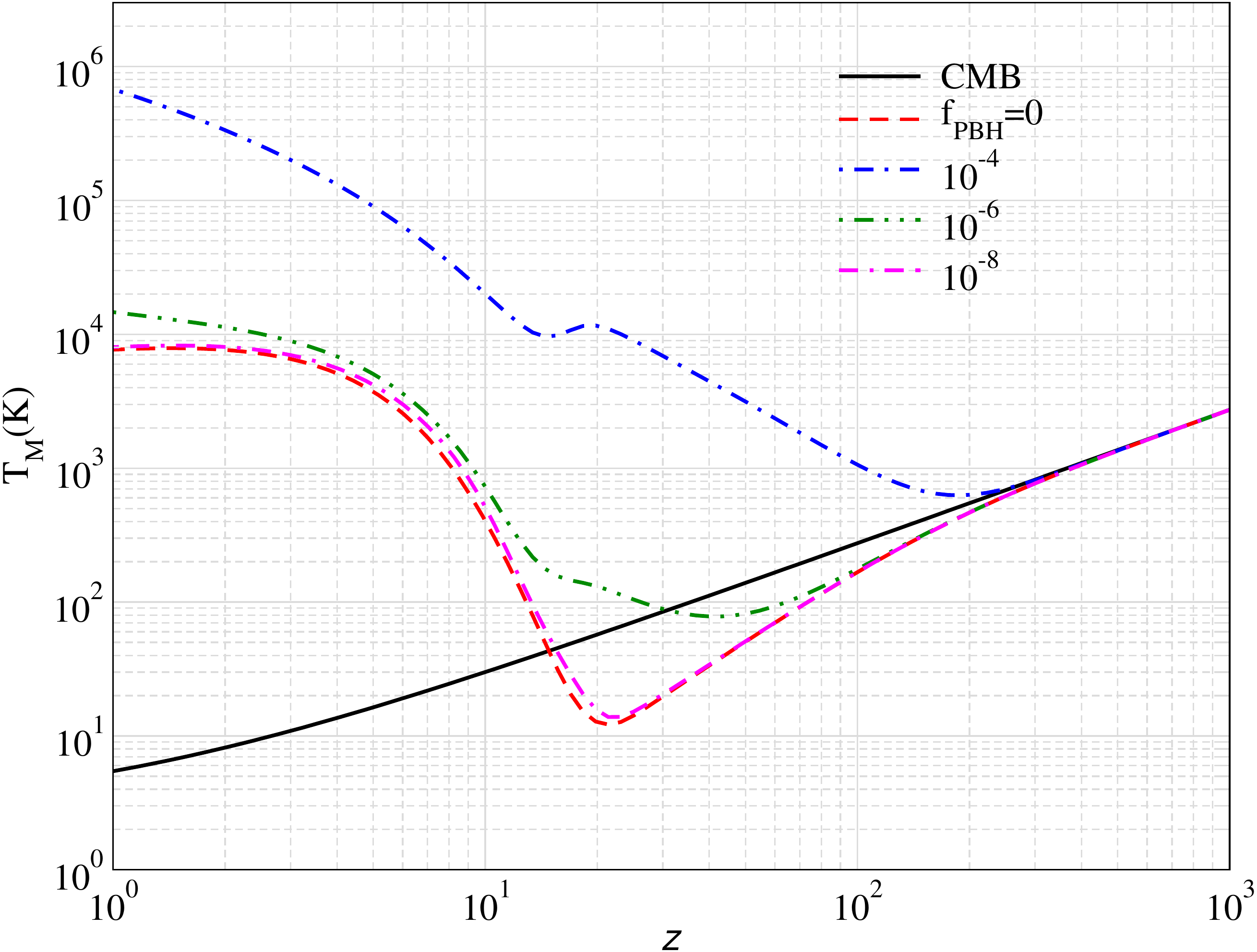}
\caption{Evolution of baryon temperature as a function of redshift for a few $f_{\rm PBH}$ as in Fig.~\ref{fig:xe}. }
\label{fig:TM}
\end{figure}


\begin{figure}
\centering 
\includegraphics[width=0.95\columnwidth]{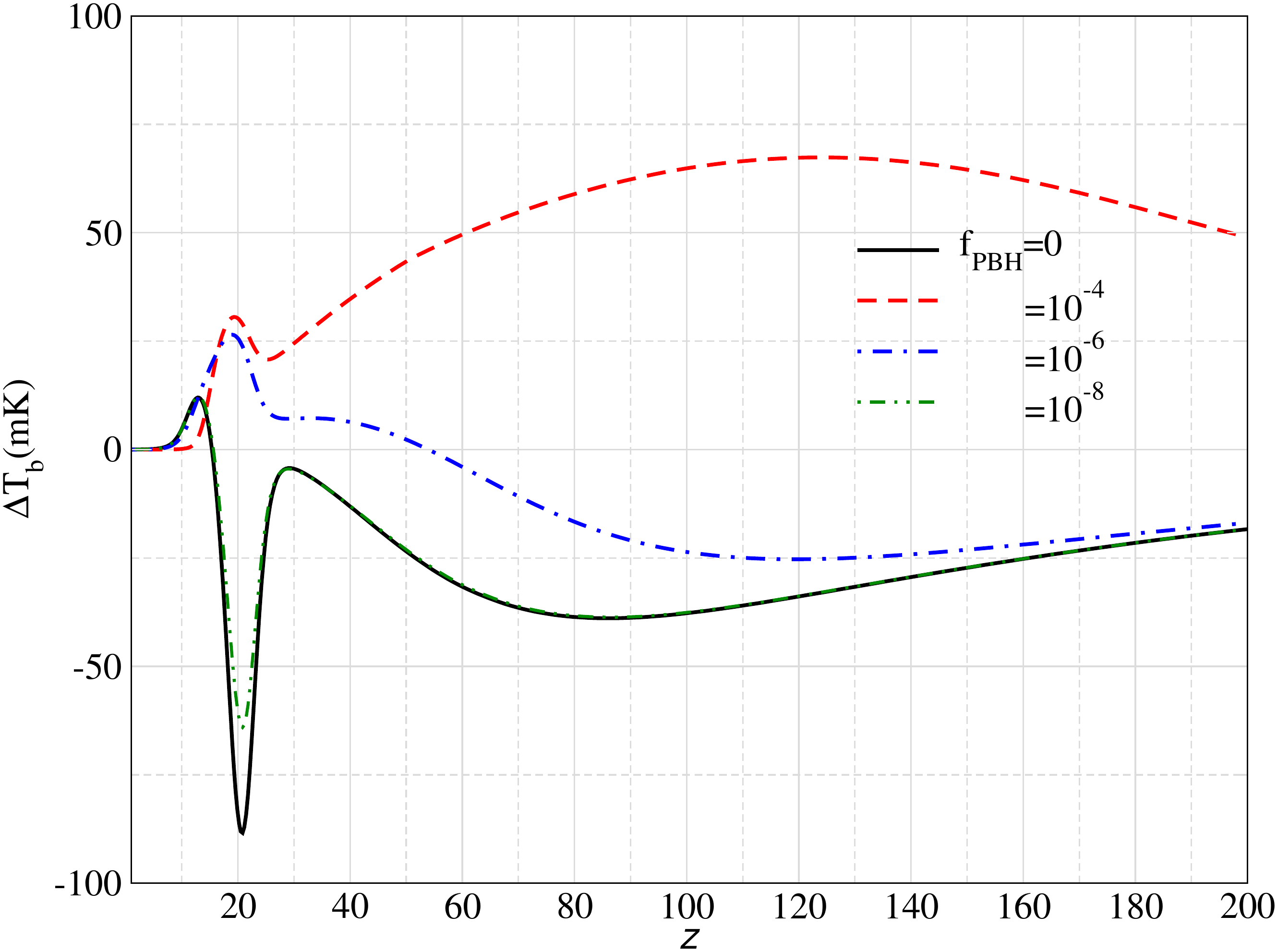}
\caption{21 cm distortion in temperature for the cases in figures above including the radio background from the accreting PBH themselves.}
\label{fig:dTb}
\end{figure}


If we assume one ionization per one photon, the ionization rate per volume is given by,
\begin{equation}
    \dot{N}(z)=\left(\frac{L_{\rm UV}}{13.6 {\rm eV}}\right) \frac{f_{\rm PBH}\,\rho_{\rm DM}}{M_{\rm PBH}}
\end{equation}
In Fig. \ref{fig:density}, we compare the number density of baryons with the ionizing photons per unit redshift bin for our chosen fiducial parameters. One can see that there are too many ionizing photons which can reionize the Universe much before $z=6$.

The change to the ionization history due to this source of photons is given by,
\begin{equation}
    \frac{\id x_{\rm e}}{\id t}\Bigg|_{\rm ion}=\frac{\dot{N}(z)}{n_{\rm H}},
\end{equation}
where $x_{\rm e}=\frac{n_{\rm HII}}{n_{\rm H}}$ and $n_{\rm H}$ is the hydrogen number density. This extra source of energetic photons modifies the recombination history of hydrogen and helium \citep{ZKS1969,P1968,SSS1999,CT2011,AH2011}. We solve for the evolution of free electrons as a function of redshift using {\tt Recfast++}. We explicitly model reionization of hydrogen and helium in this calculation. In Fig. \ref{fig:xe}, we plot the evolution of free electron fraction as a function of redshift for few different $f_{\rm PBH}$. Even for $f_{\rm PBH}=10^{-6}$, there is a significant build up of free electrons before $z=10$ which may violate CMB anisotropy constraint.
For $f_{\rm PBH}=10^{-4}$ some of the simplifying assumptions of the recombination calculation may also be violated; however, this regime seems to be ruled out either way.

 In Fig.~\ref{fig:TM}, we plot the temperature of baryonic gas for  $f_{\rm PBH}$ as shown in Fig.~\ref{fig:xe} using {\tt CosmoTherm} \citep{Chluba2011therm}. There is big deviation of gas temperature from CMB at $z>100$ for $f_{\rm PBH}=10^{-4}$ which is expected to wash out the 21cm absorption signal. For smaller $f_{\rm PBH}$, adiabatic cooling plays an important part and gas temperature falls below CMB temperature. We also plot the 21 cm distortion temperature as a function of redshift in Fig.~\ref{fig:dTb}. For $f_{\rm PBH}=10^{-4}$ and even for $10^{-6}$, the absorption feature is completely washed out at $z\approx 10-20$. As mentioned in Sect. \ref{sec:21cmmodel}, the WF coupling could be induced earlier due the Ly$\alpha$ contribution from the PBHs, which has been neglected in our calculations. This would not have any effect at redshifts $z>100$ since collisional coupling to $T_M$ is strong enough. For lower redshifts, the effect in case of $f_{\rm PBH}=10^{-4}$ would be to further strengthen the emission signal. In the case of $f_{\rm PBH}=10^{-6}$, the absorption signal in $z\approx 50-100$ range will be more pronounced, but there would still be an emission signal at  $z\simeq20$. We can obtain a conservative constraint on $f_{\rm PBH}$ by requiring that we see 21 cm signal in absorption (i.e $T_M<T_{\rm CMB})$ at $z\simeq 20$. In that case, constraints on $f_{\rm PBH}$ turns out to be $\lesssim 10^{-6}$. This number can be further tightened by using detailed astrophysical modelling. Therefore, we see that $f_{\rm PBH}=10^{-4}$, as used in \cite{MK2021} to explain the radio excess today, may run into severe constraints when the UV photon emission from PBHs are taken into account.

\subsection{Limits from CMB anisotropies}
\label{sec:CMB_aniso}
Changes to the ionization history affect the CMB temperature and polarization anisotropies \citep{Peebles2000, Chen2004}. Since the latter have been accurately measured using \Planck \citep{Planck2013power, Planck2015params, Planck2018params}, one can directly convert the changes to the ionization history into a limit on $f_{\rm PBH}$ by using an $\xe$ principal component projection method \citep{Farhang2011, Farhang2013, Hart2020PCA}. 

We refer the reader to \cite{Hart2020PCA} for details of the projection method, but in brief, given the PBH model parameters, we can compute $\xi(z)=\Delta \xe/\xe$ with respect to the standard ionization history. The $\xi(z)$ response can then be projected onto the first three $\xe$-modes, $E_i(z)$, to obtain the relevant mode amplitudes\footnote{For $f_{\rm PBH}=10^{-6}$ we obtain $\mu_1\approx 0.051$, $\mu_2\approx -0.026$ and $\mu_3\approx 0.170$.} $\mu_i=\int E_i(z)\,\xi(z)\,{\rm d}z$. Assuming that the responses in the $\mu_i$ are linear\footnote{We confirmed this statment for $f_{\rm PBH}\lesssim \pot{\text{few}}{-6}$.} in $f_{\rm PBH}$ and using the \Planck constraints from \cite{Hart2020PCA}, we then find $f_{\rm PBH}\lesssim \pot{3}{-6}$ (95\% c.l.). This limit falls into a similar regime as the one obtained in \cite{AK2017} for PBHs with $M\lesssim 10^4 M_{\odot}$. Performing a simple extrapolation of the constraint contour in Fig.~14 of \cite{AK2017} to $M=10^5 M_{\odot}$, we see that the weakest constraint from collisional ionization is also of the order of $f_{\rm PBH}\approx 10^{-6}$. Therefore, our calculations rules out the model proposed in \cite{Mittal2022}.

We should point out that we actually constrain the total UV luminosity from black holes, which is proportional to the product $f_{\rm duty}f_X\lambda f_{\rm PBH}$. Only when we choose a particular value of astrophysical parameters, we obtain a constraint on $f_{\rm PBH}$. 
We remind the reader that \cite{MK2021} fixed $f_{\rm PBH}$ at $10^{-4}$ and boosted the combination of $f_{\rm duty},f_X\lambda, f_{\rm PBH}$ by a factor of 20 to explain the ARCADE excess. According to the parametric relation used in \cite{MK2021} and in this work, the radio emissivity is proportional to X-ray luminosity, which in turn is proportional UV luminosity. Therefore, if we fix $f_{\rm PBH}$ and choose the value of astrophysical parameters to tune radio emissivity to match the radio excess observation we still run into strong CMB anisotropy constraints. 

In these calculations we have assumed that the UV photons escape to the intergalactic medium (IGM) and lead to uniform ionization and heating of the universe. If instead we are able to trap these photons locally, close to black holes, it may be possible to avoid CMB anisotropies constraints. The most energetic X-ray photons can still escape to the IGM and propagate large distance before ionizing and heating the medium. We plan to perform a detailed calculation in a future paper.
Alternatively, the PBH luminosity may not follow the parametric relation used in this calculation and may be radio-loud by at least a factor of hundred. This may be a possible explanation for ARCADE and EDGES excess without running in the CMB anisotropy constraints. However, there are stringent constraints on the PBH abundance from CMB spectral distortions, as we show now.


\subsection{Limits from \COBEF}

\begin{figure}
\centering 
\includegraphics[width=0.95\columnwidth]{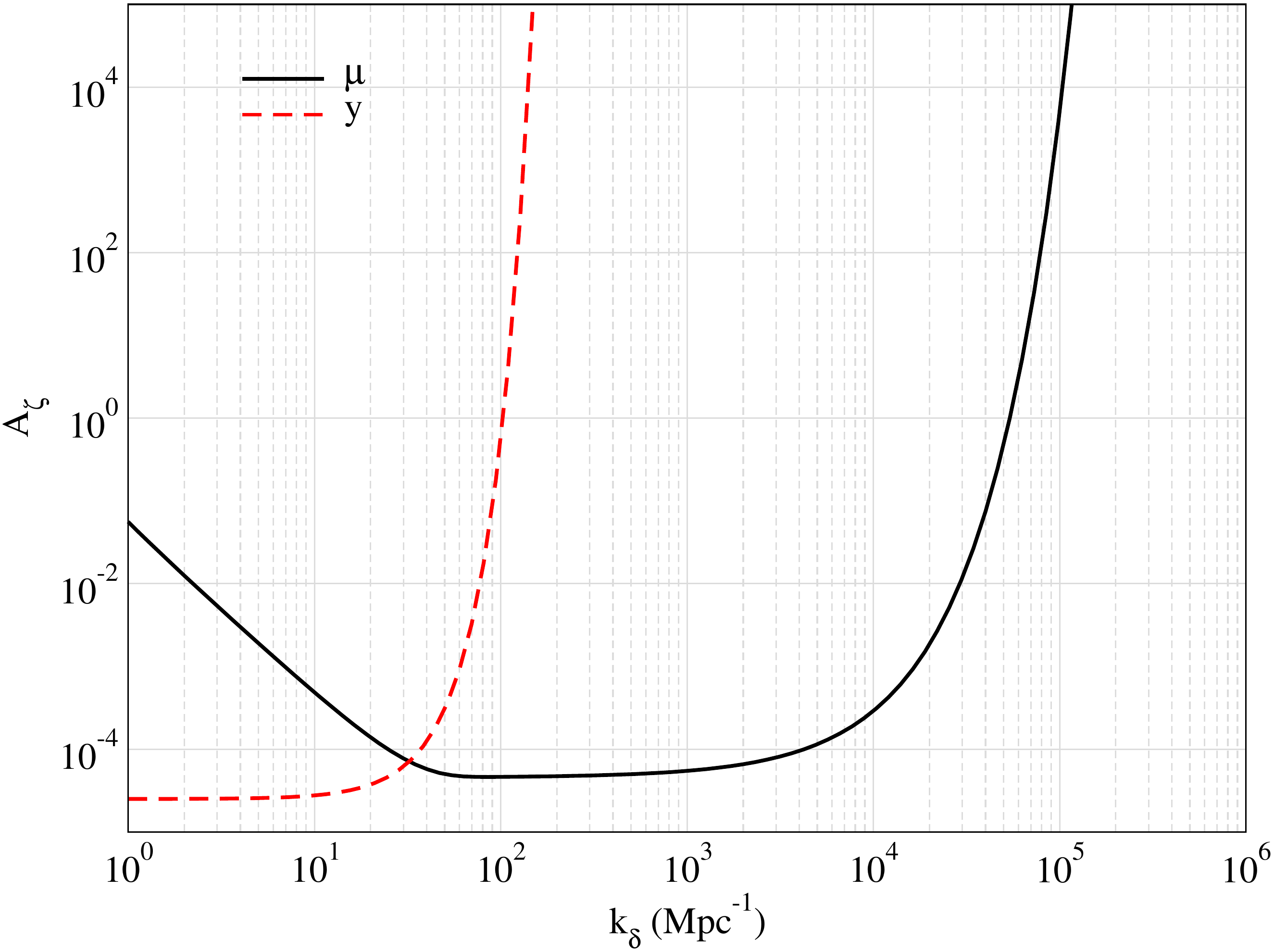}
\caption{Allowed $A_{\zeta}$ from $y$ and $\mu$-distortion constraint \citep{Fixsen1996}.}
\label{fig:A_zeta}
\end{figure}


\begin{figure}
\centering 
\includegraphics[width=0.95\columnwidth]{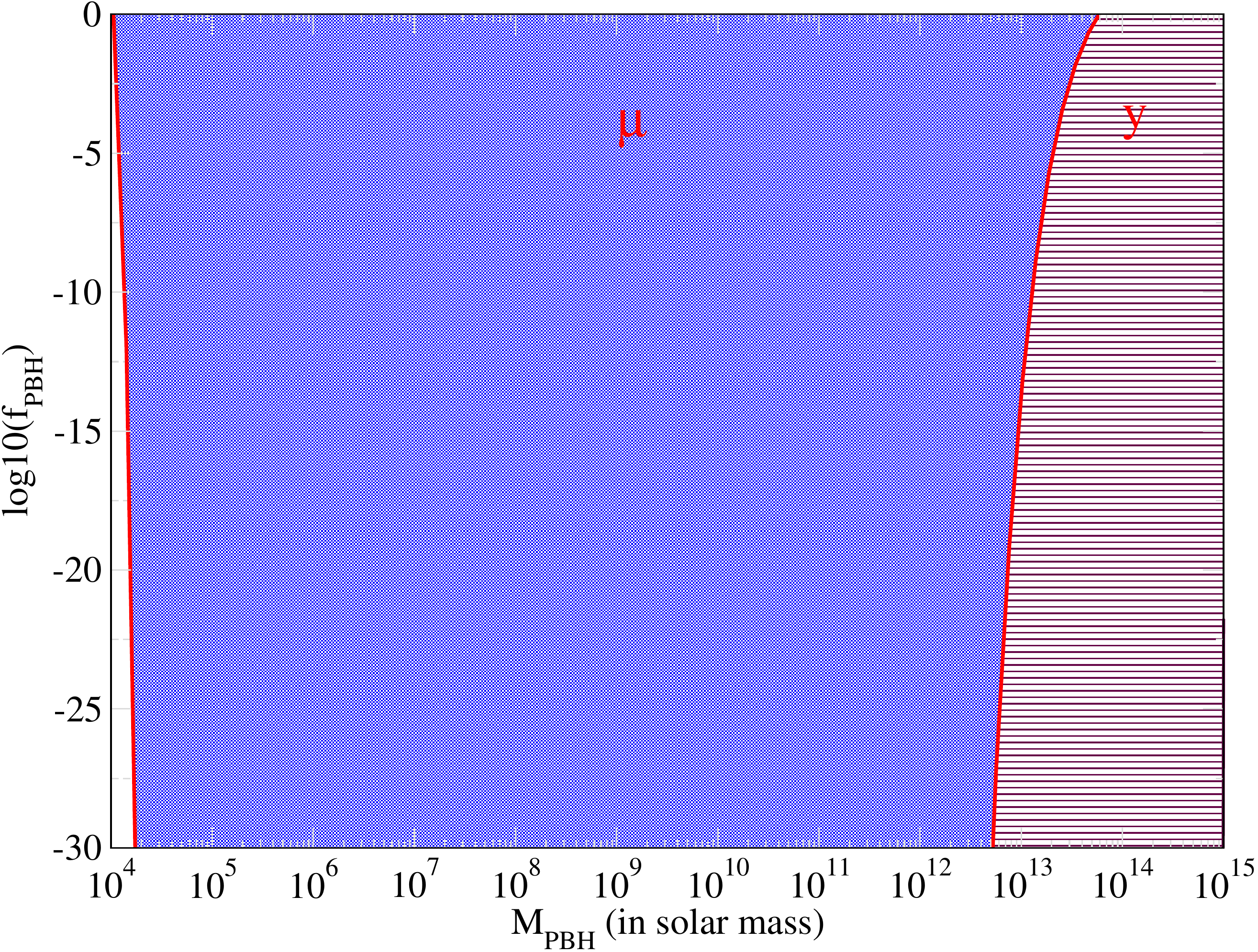}
\caption{Constraint on $f_{\rm PBH}$ from $y$ and $\mu$-distortion constraints \citep{Fixsen1996} as a function of PBH mass. The excluded region is shown in filled lines.}
\label{fig:fbh_cobe}
\end{figure}


Primordial black holes are formed from density fluctuations in the early universe \citep{Carr1975}. These same perturbations also give rise to CMB temperature fluctuations. With the evolution of the universe, the CMB blackbodies with slightly different temperature mix with each other. This gives rise to CMB spectral distortions as the sum of two black bodies is not necessarily another blackbody subject to constraint on photon non-conserving processes \citep{Sunyaev1970mu, Daly1991, Hu1994, Chluba2012, Khatri2012short2x2}. Therefore, CMB spectral distortion constraints \citep{Fixsen1996} can indirectly constrain the abundance of supermassivee PBHs in the mass range of $\approx 10^4-10^{12} M_{\odot}$ \citep{Chluba2012inflaton,KNS2014}. Here, we give a brief overview of the calculations of constraints for gaussian initial conditions. The readers can find more details in \cite{Chluba2012inflaton,KNS2014}. The curvature power spectrum is given by,
\begin{equation}
    P_{\zeta}(k)=2\pi^2 A_{\zeta}^{\delta}k^{-2}\delta(k-k_{\delta})
\end{equation}
We assume very sharp density fluctuations at scale $k_{\delta}$ which gives rise to monochromatic distribution of PBHs.\footnote{In \citet{Chluba2012inflaton}, Fig.~2, a more detailed window function was used to convert to a limit on $A_{\zeta}$; however, the number are comparable.} For a single $k$ mode, $\mu$-distortion generated is given by \citep{Chluba2012inflaton},
\begin{equation}
    \mu\approx 2.2A_{\zeta}\left[\rm{exp}\left(-\frac{k_{\delta}}{5400}\right)-\rm{exp}\left(-\left[\frac{k_{\delta}}{31.6}\right]^2\right)\right].
\end{equation}
Similarly, for $y$-distortion, one has \citep{Chluba2012inflaton},
\begin{equation}
    y\approx 0.4A_{\zeta}\rm{exp}\left(-\left[\frac{k_{\delta}}{31.6}\right]^2\right)
\end{equation}
In Fig.~\ref{fig:A_zeta}, we plot the constraint on $A_{\zeta}$ assuming generated distortions to be of the order $\mu\approx 10^{-4}$ \citep{Fixsen1996} and $y\approx 10^{-5}$.

The probability of having a collapsed objects at a density threshold $\delta_{\rm th}$ is given by \citep{KNS2014},
\begin{equation}
    \beta\approx 2{\rm exp}\left(-\frac{{\rm e}^2\delta_{\rm th}^2}{8P_{\delta}}\right),
\end{equation}
with $\rm e=2.718$, $\delta_{\rm th}=0.5$ and $P_{\delta}=\frac{16}{81}A_{\zeta}$ \citep{KNS2014}. The abundance can be related to $f_{\rm PBH}$ \citep{NSK2017} as,
\begin{equation}
    \beta\approx1.1\times 10^{-8}\gamma^{-1/2}\left(\frac{g}{10.75}\right)^{1/4}\left(\frac{\Omega_{\rm DM}}{0.27}\right)^{-1}\left(\frac{M}{30M_{\odot}}\right)^{1/2}f_{\rm PBH},
\end{equation}
where $\gamma\simeq O(1)$ and $g$ is the number of degrees of freedom of relativistic particles.

We show the constraints on the abundance of PBHs from CMB spectral distortions in Fig.~\ref{fig:fbh_cobe}. We see that the $\mu$-distortion constraint essentially excludes any PBHs in the mass range of $10^4-10^{12}M_{\odot}$. The large density fluctuation forms in the exponential tail of the assumed gaussian initial power spectrum. Due to the strong $\mu$-distortion limit the width of gaussian is reduced, which renders high density fluctuations extremely unlikely.

The constraints are model-dependent and do vary if the initial conditions of the universe are non-gaussian. In the case, the probability for PBH formation at a given level of the curvature perturbation can be significantly enhanced over the gaussian case.
Depending on these details one can evade the CMB spectral distortion constraints \citep{NSY2016} or obtain constraints which are orders of magnitude stronger than the CMB anisotropy limits obtained in this paper \citep{NCS2018}. There is even a possibility to start with smaller mass PBHs ($\lesssim 10^4 M_{\odot}$) which can then accrete and become supermassive at the redshifts that we are interested in. 
This regime of PBH masses could be directly constrained by future CMB spectral distortion measurements \citep{Chluba2021ExA}, which promise improved limits on $\mu$ by many orders of magnitude.

We also mention the direct $y$-distortion constraint. The late heating of the Universe by the $X$-ray and UV photons that lead to early reionization will exceed the \COBEF limit of $y\lesssim \pot{1.5}{-5}$ (95\% c.l.) once $f_{\rm PBH}\gtrsim 3\times 10^{-4}$. In a similar way as the $\mu$-distortion, the $y$-distortion limit can furthermore be used to place a constraint on $A_\zeta$ \citep{Chluba2012inflaton}. Looking at Fig.~\ref{fig:A_zeta}, we find that one can obtain a stronger limit for PBHs with mass $\gtrsim$ $10^{13}M_{\odot}$ from $y$-distortion. These constraints are orders of magnitude stronger than other constraints in this mass range \citep{CKV2021}. Aside from the fact that we have never encountered BHs of this size, this is reassuring.


\section{Discussion and conclusions}
\label{sec:conclusions}

In this paper, we have used the empirical relations between radio, X-ray and UV luminosities to show that it is unlikely that accretion on PBHs is able to explain the observed radio excess seen by \cite{Arcade2011} and \cite{DT2018}. The required abundance of PBHs that can explain the radio excess results in the emission of too many ionizing photons which can ionize the universe much before reionization, which can be ruled out by \Planck (see Sec.\ref{sec:CMB_aniso}). Ionization of neutral hydrogen also results in gas heating which can raise the temperature of gas to $\approx 10^4$K. We obtain strong CMB anisotropy and spectral distortions constraints on PBH abundance which are much lower than the assumed $f_{\rm PBH}=10^{-4}$ in \cite{MK2021}, which would reduce the radio emissivity and the corresponding radio background as seen today. 

Previously the authors in \cite{HHMRVV2018,MPVW2019,Y2021} have obtained constraints on accreting PBHs by studying their effect on 21cm signal in the mass range $M\lesssim 10^4 M_{\odot}$. They computed the luminosity of black holes assuming a theoretical model of accretion onto the black holes. However, we do not use any such model but use the empirical relation between luminosity at different frequency bands. This relation is assumed to hold at all redshifts.
Radio emission from accreting astrophysical black holes, their implication in the context of EDGES result \citep{Edges2018} and impact of UV and X-ray emission on reionization was carried out in \cite{ECLDSM2018} [see also \cite{ECL2020}]. As opposed to primordial black holes, astrophysical black holes form only at $z\lesssim 30$ and accretion was assumed to stop by $z\simeq 15$, in their work. 

In comparison, the primordial black holes can accrete over a much broader range of redshifts. Also in the case of astrophysical black holes, it is assumed that only a fraction ($f_{\rm BH,esc}$) of ionizing photons is able to escape to intergalactic medium which can then ionize neutral hydrogen. Typically the value of $f_{\rm BH,esc}$ is within 0.01-0.1 but it can be highly uncertain \citep{MKHFQKM2015}. This effectively reduces the UV luminosity of astrophysical black holes. However, no such criteria exists for primordial black holes and no such factors of $f_{\rm BH,esc}$ are assumed  while deriving the constraints on PBH abundance in the aforementioned works. Therefore, we expect to find stronger constraints on the abundance of PBHs as compared to astrophysical black holes for same intrinsic UV luminosity.   

While we obtain a conservative estimate on abundance of supermassive PBHs -- the primary motivation for our work -- one can obtain more accurate constraints on $f_{\rm PBH}$ with detailed calculations of ionization history and 21 cm signal at $z\simeq 20$ including astrophysical Ly$\alpha$ and X-ray modelling. We defer this work to the future, expecting that the general conclusion remains. 

Since, as we argued, for extremely radio-loud PBHs it may still be possible to evade the CMB anisotropy constraint, it would be very important to study the physics of the atmospheres of the PBHs in more detail. In addition, future CMB spectral distortion limits in combination with astrophysical limits could help close existing loopholes that could lead to an early formation of super-massive PBHs at $z\gtrsim 100-200$.

{\small
\section*{Acknowledgments}

SKA would like to thank Shikhar Mittal, Girish Kulkarni and Rishi Khatri for discussions.
This work was supported by the ERC Consolidator Grant {\it CMBSPEC} (No.~725456).
JC was furthermore supported by the Royal Society as a Royal Society University Research Fellow at the University of Manchester, UK (No.~URF/R/191023).
}

\section{Data availability}
The data underlying in this article are available in this article.

{\small
\vspace{-3mm}
\bibliographystyle{mn2e}
\bibliography{Lit}
}
\newpage

\end{document}